**Processing of dense bio-inspired ceramics with deliberate microtexture**

Hortense Le Ferrand[1], Florian Bouville[1,*], André R. Studart [1,*]

[1] Complex Materials, Department of Materials, ETH Zürich, Switzerland

*florian.bouville@mat.ethz.ch, *andre.studart@mat.ethz.ch

**Abstract**

The architectures of biological hard materials reveal finely tailored complex assemblies of mineral crystals. Numerous recent studies associate the design of these local assemblies with impressive macroscopic response. Reproducing such exquisite control in technical ceramics conflicts with commonly used processing methods. Here, we circumvent this issue by combining the recently developed Magnetically-Assisted Slip Casting (MASC) technique with the well-established process of Templated Grain Growth (TGG). MASC enables the local control over the orientation of platelets dispersed among smaller isotropic particles. After a high temperature pressure-less treatment, the grains of the final ceramic follow the same orientation of the initial platelets. This combination allows us to produce 95 % dense alumina part with a grain orientation following any deliberate orientation. We successfully fabricated microstructures inspired from biological materials with ceramics that present periodically varying patterns with a programmable pitch of a few tens of microns. We confirmed the capacity of the process to tailor local mechanical properties through local grains orientation using micro-indentation. This micrometer scale control over the local mechanical properties could be applied to adapt ceramic structures to complex loads using this inexpensive and scalable process. In systems where functional properties also depend on anisotropic grain orientation, the principle presented here could enable the creation of new multifunctional ceramics.

## Introduction

Ceramics are ubiquitous in extreme environments due to their ability to sustain harsh chemicals and high temperatures while being mechanically loaded. As for most polycrystalline materials, the mechanical response of dense ceramics is strongly correlated to their internal grain structure (*1, 2*). Generally, ceramics with submicron sized grains are harder and stronger but less tough than ceramics with larger grains (*3–6*). In addition, the presence of anisotropy in the form of elongated inclusions or grains has been found to enable crack deflection mechanisms and therefore to increase the energy needed to propagate a crack through the material (*7, 8*). Functional properties coming from the crystalline lattice, for instance piezoelectricity, also depends strongly on grain orientation (*9*). Controlling the size, shape and orientation of the grains is thus a prerequisite for the fabrication of dense ceramics adapted to their specific use. Traditionally, this is achieved by optimizing the high temperature sintering step that transforms an assembly of particles into a dense part through mass transport. The resulting particle's rearrangement and change in size depend highly on the particle's chemical composition, homogeneity, size, aspect ratio and global packing, along with the heating rate, atmosphere and the final sintering temperature. Without deliberately promoting the formation of oriented microstructures, the ceramics obtained through conventional sintering will present isotropic grains or randomly oriented elongated crystals. To fabricate a dense ceramic with anisotropic mechanical response or local variations in hardness, strength or toughness, it is thus required to purposely control the orientation of the grains within the sintered ceramic structure.

Deliberately textured dense ceramics can be obtained by methods that combine control over particle orientation and densification. The term texture refers here to a preferred orientation of grains in a particular direction in the material. Spherical particles with anisotropic magnetic susceptibility in their crystallographic structure can be oriented during casting using ultra high magnetic fields of 10 T, resulting in a textured ceramic after sintering (*10*). To avoid the use of such powerful magnets, anisotropic particles have been used as starting materials and aligned by tape-casting (*9*), electric fields (*11, 12*) or via ice-templating followed or not by hot pressing (*13–15*). Another approach is to use the preferential crystallographic orientation of these anisotropic templates to grow oriented larger grain structures during the sintering, a method called templated grain growth (TGG) (*16*). Despite how complex these synthetic microstructures appear to be, they pale in comparison to natural ones. One particularly interesting natural structure, called Bouligand structure, consists of layers of aligned mineral needles that are progressively rotated with respect to their neighbor, forming a periodic pattern. This assembly is recurrently found in the endoskeletons and exoskeletons of mammals and invertebrates (*17, 18*) and is closely related to outstanding mechanical performance with unprecedented impact resistance (*19–21*). The fabrication of such exquisite variation in local texture has been recently achieved in composite materials up to 50 vol% mineral content using additive manufacturing methods based on 3D printing (*22–25*) or magnetic alignment with slip-casting (MASC) (*26*).

These latest efforts to bring complex heterogeneous texture into polymer/ceramic composites have not yet reached dense ceramics. The major hurdle is the necessary uniaxial pressing step used for densification that inevitably erases the microstructure. A recent study successfully oriented large anisotropic alumina microplatelets in a suspension of small alumina nanoparticles using low magnetic fields, but the gelling reaction used to maintain the particles in place does not allow for a local control of their orientation (*27*). Tailoring the texture at the micrometer scale to explore the effect of local tuning of the properties therefore calls for new processing strategies.

In this study, we present a scalable and programmable bottom-up method to fabricate 95% dense ceramics with unprecedented microtexture resulting in unusual control over the local mechanical properties. The technique combines magnetically-assisted slip casting (MASC) to enable the fabrication of periodic microstructures with sub-millimeter pitches, with pressure-less densification by templated grain growth (TGG). The process can be computer-controlled to fabricate periodic textures, offers possibilities of 3D shaping and flexibility in final grain size. Building up periodic microstructures inspired from biological hard materials could provide avenues for technical ceramics with locally tailored response or impact-resistance properties. Beyond mechanics, the process can be extended to locally tune other grain-size dependent functions such as thermal conductivity, optical properties or piezoelectric response.

# Processing path for the densification of alumina ceramics with controlled texture

A combination of two processes, magnetically-assisted slip-casting and templated grain growth, was used for the densification of alumina ceramics with controlled texture (Fig. 1). Programmed microstructures of anisotropic particles are enabled by the first process (*26*), whereas the second allows to grow grains with crystallographic orientation and shape that follow the one of the anisotropic particles (*16*).

Figure 1A describes the five steps of the process. Prior to assembly, anisotropic alumina microplatelets, the template particles, of 8 μm diameter on average, are functionalized with 0.1 vol% of superparamagnetic iron oxide nanoparticles. These decorated particles then exhibit ultra-high magnetic response (*28*) and can be manipulated using magnetic field as low as a few millitesla. Additionally, this very low amount of surface modification maintains the overall positive surface charge of the platelets when dispersed in acidic pH (*27*), and therefore enables efficient de-agglomeration with positively charged nanoparticles at high solid loading using a rotary ball-mill. The colloidal suspension used for the assembly contains 50 vol% solid with magnetized platelets and alumina nanoparticles of 180 nm diameter at a ratio 10:90. In the liquid suspension obtained after dispersion, the magnetized platelets align under the rotating magnetic field generated by a magnet rotating at a frequency of 1 to 3 Hz and with approximately 50 mT strength at the area of the alignment (*29*). When this suspension is casted onto a porous mold, typically made of gypsum, the suction of the liquid through the pores locally increases the concentration of particles at the surface of the mold. This phenomenon jams the particles together, both platelets and nanoparticles, in their position and thus fixes the orientation provided by the magnetic field. After complete drying of the assembled powders, the consolidated body can be removed from the mold and sintered at 1600°C for 1h. This yields a 95 ± 1% dense ceramic with a microstructure where elongated grains have the same orientation as the magnetically-oriented templates.

Since powder sintering is driven by the diminution of the particle surface, the smaller particle with a larger specific surface will present faster sintering kinetics than the large template. Consequently, the atoms from the small particles will diffuse at high temperature along the crystallographic planes of the larger particles, increasing the templates' dimensions at the expense of the small particles (*11*). The diffusion is further facilitated by the presence of doping atoms at the surface of the large particles, in this case iron (*30*). However, the template grain growth reflects the expected driving force for grain growth where the smaller dimensions of the templates diffuse and sinter faster then their length. This difference in kinetics translates into a higher growth of the grain along the thickness of the templates, with nearly 570% increase in size in the thickness compared with only 25% in the length. Indeed, in the conditions selected for this study, the final alumina grains obtained have an aspect ratio of 5 with an average diameter of 10 μm and thickness of 2 μm. Using liquid phase sintering by the addition of other dopants like silica, calcium, magnesium or titanium could be used to further vary the grain sizes and aspect ratio and final density through liquid phase sintering (*16*).

Using this bottom-up approach based on MASC and TGG, we can produce dense alumina ceramics with a texture directly resulting from the magnetic alignment of the templates. Since no pressing is required for the densification, it is possible to exploit the full potential of magnetic manipulation: its ability to orient anisotropic platelets at along any direction in space.

**Texture orientation at any deliberate angle**

Applying rotating magnetic fields with a well-defined direction of alignment, 95% dense and textured alumina ceramics were fabricated with any desired texture orientation (Fig. 2). This texture direction is defined in the following by the angle $\phi$ between the normal to the basal plane of the grain and the horizontal.

The electron micrographs of fractured surfaces of the ceramics obtained using template alignment directions at various angles $\phi$ confirm the development of the texture in so far unachievable directions (Fig. 2A). The 2D-Small-Angle X-ray Scattering (2D-SAXS) pattern presented in Figure 2B was obtained on an area of 20 μm per 20 μm on a sample with vertically aligned grains as pictured in the electron micrograph $\phi=0°$ (Fig. 2B). It exhibits sharp and intense interference maxima at ± 9° from the horizontal and two less intense interference maxima at ± 64° and ± 52°. Interference patterns appear at 90° from the orientation of the physical object they originate from, so the primary maxima of the pattern can be attributed to scattering of vertically aligned entities. Since SAXS probes features between 2 and 300 nm, it is probable that it is the interfaces along the long axis of the grains or the elongated pores located there that are causing the scattering. With similar reasoning, it is also probable that the two other maxima come from the short edges of the grains. These short edges typically have angles of around 45° with respect to the grain long axis, as highlighted in yellow in the micrograph in figure 2B.

The surface of the templates obtained by molten salt synthesis corresponds to the crystallographic basal plane of miller indices (00l). After the sintering, the grains obtained also exhibit a corresponding preferred crystallographic orientation along this group of plane, the most intense one being the (0012) measured by X-Ray diffraction (Fig. S1). The position of the (0012) reflections on a polished cross-section of a sintered sample was also used to quantify the grain degree of misalignment. During these experiments, the sample is lying flat on a stage while the incident X-rays are in theta-2theta position to obtain the (0012) plane reflection, at a 2-theta angle of 90.7° for $\alpha$-alumina. The sample is then rocked around its axis by an angle $-40° < \omega < 40°$ while the reflected intensity is recorded. The full width at half maximum (FWHM) of the rocking curve is thus a direct measure of the grain orientation quality (Fig. 2C) (*31*). Grain misalignment measured by image analysis of scanning electron micrographs gives similar FWHM and therefore was used for the characterization of the misalignment at angles $\phi$ inaccessible to a XRD setup, as well as for images from the literature (Fig. 2D). The misalignment in horizontally textured dense ceramics ($\phi=90°$) is 10°, a value similar to the one obtained by other methods such as tape-casting (*11*), layer-by-layer deposition (*32*) or magnetic alignment in polymer (*33*) and lie within the range of the misalignment of the nacreous layer of seashells (Fig. S2), the closest biological material in terms of microstructure from our sample. Ice-templating followed by pressure-assisted sintering can also produce dense and textured ceramics, but with a higher misalignment, in the range of 20° (*13*). However, the uniqueness of the MASC-TGG method presented here is its ability to provide ceramic materials with similarly high degree of alignment control but at any given angles. To illustrate this, we measured the degree of misalignment in ceramics with a texture oriented at angles $\phi$ of 0, 13, 45 and 78°. In these specimens, the grain misalignment is comprised between 15 and 25°, the lower value obtained for $\phi = 0°$. The lower alignment obtained for $\phi$ equals to 13, 45, and 78° could arise from both gravity or from the flow of liquid during the drying and could be improved by further optimizing the casting conditions (*34*).

Bulk ceramics with precise control of their texture direction adds tailored anisotropy to the traditional TGG process. In addition, the jamming front position is moving inside the suspension during the slip-casting by MASC. Taking advantage of this particularity, periodic microstructures in porous particle assemblies have been obtained by varying the rotating magnetic field orientation during the casting (*26*). Applying this principle to our suspension will thus enable the fabrication of dense ceramics with locally controlled texture.

**Fabrication of periodically textured ceramic**

During the slip casting, a step-wise rotation of the magnetic alignment direction with respect to the sample produces a colloidal assembly with periodic orientations of the template particles. After sintering, this assembly converts into a dense ceramic with a periodic orientation of its constitutive grains. As a proof-of-concept, we prepared ceramics with vertical orientations of the grains, rotating in the perpendicular direction at chosen angles (Fig. 3).

The setup used to produce these periodic assemblies consists in a permanent magnet rotating around a horizontal axis, and a porous casting mold that will receive the sample placed below the magnet, on a stepper motor. The magnetic field produced by the rotating magnet will align the template particles vertically ($\phi = 0°$) while the stepper motor will rotate the sample in-plane by a chosen angle $\theta_R$ at every defined time interval $\tau$ (Fig. 3A). The gypsum support of the sample will slowly remove water and concentrate the particles until they jam, fixing the templates' orientation. During a rotation of the stepper motor of an angle $\theta_R$, the templates above the jamming front stay within the alignment plane defined by the rotating magnet while the jammed templates rotate by the same angle $\theta_R$. The sequence of sample rotations as the jamming front grows in the sample produces eventually a periodic layered structure. In this particular case, it reproduces the Bouligand structure (Fig. 3B).

After the sintering at 1600°C, during which the templated grain growth occurs, the sample presents a grain orientation that periodically varies along the z-direction. This periodic change in grain orientation is revealed on the sample polished cross-section (Fig. 3C). The brighter bands on the images correspond to layers where the basal planes of the grains are facing the observer, whereas the darker bands correspond to layers where the short edges of the grains are directed towards the observer. As the process is based on slip-casting, the sample's macroscopic shape is directly the imprint of the mold, giving a large degree of freedom in the final morphology. More importantly, the shape of the jamming front in the suspension is also related to the mold surface topography, which translates in our process into the conformation of the textured grains layers with the mold surface as well (Fig. 3D). In addition, the overall size of the casted sample is solely depending on the size of the mold and on the area where the magnetic field is strong enough to orient the templates. In this study, we fabricated discs of 25 mm diameter and 5 mm in thickness using only a 100 mm² neodymium magnet.

A closer investigation of the microstructure with a SEM allows the appreciation of the grains alignment within each period (Fig. 3E). In opposition with the bilayer structures containing only one perpendicular rotation of vertical platelets that morph during sintering due to perpendicular directions of shrinkage (*27*), the ceramics prepared here did not exhibit any relevant shrinkage nor noticeable macroscopic distortions after sintering. The dimensions of each layer, below 100 µm, and the total thickness of the sample, roughly 5 mm, should account for the inability of one layer to stress the whole structure. As a result, no crack or wrapping developed. However, the local tilting of the platelets observed in the SEM image in figure 3E could be the result from these internal stresses.

Furthermore, when the jammed layer increases in thickness during the casting, it opposes a larger pressure drop upon water removal. As a consequence, the jamming front speed decreases with time, and directly affects the pitch of the periodic structures in a predictable fashion. Indeed, the jamming front line at a position *z* is expressed by:

$$t(z) = Az^2, \text{ with } A = \frac{\eta R}{2J \Delta P}, \qquad \text{Eq. 1}$$

where $\eta$ is the viscosity of the liquid, $R$ is the hydrostatic resistance of the cast layer and $J$ is the ratio between the volume of cast layer and of the extracted liquid (*26*). In this system, A=119 s.mm$^{-2}$ (Fig. S3). In this case, the dependence of the platelets alignment direction around the z-axis, defined as the angle $\theta$, with time *t*, is described by the following equation:

$$\theta(t) = \sum_{m=1}^{+\infty} \theta_R \, H(t - m\tau), m \in \mathbb{N} \qquad \text{Eq. 2}$$

where *H* is the Heaviside step function and *m* an integer. For this function, the period T of the temporal variation of the angle $\theta$ is given by $\pi\tau/\theta_R$. Combining Eq. 1 and Eq. 2 gives the relation for the local texture orientation as a function of the position *z* from the mold surface:

$$\theta(z) = \sum_{m=1}^{\infty} \theta_R \, H(z - \sqrt{\frac{m\tau}{A}}), m \in \mathbb{N} \qquad \text{Eq. 3}$$

This implies that the $n^{th}$ complete rotation of the templates from $\theta = 0$ to $\theta = \pi$, which corresponds to one pitch, will occur at distances $z_n$ from the surface given by:

$$z_n = \sqrt{m\tau/A} \text{ for } m \text{ values equal to } n\pi/\theta_R \qquad \text{Eq. 4}$$

Consequently, if the time interval $\tau$ between each rotation is maintained constant, the pitch $p = z_n\text{-}z_{n-1}$ of the structure decreases.

Furthermore, the pitch $p$ of a microstructure with $\theta_R = 90°$ is expressed by the following equation as demonstrated in ref (*26*):

$$p_{90} = \frac{\tau}{2Az} \qquad \text{Eq. 5}$$

then for $\theta_R < 90°$, the pitch of the microstructure can be expressed directly as a function of this reference pitch $p_{90}$:

$$p_{\theta_R} = \frac{90}{\theta_R} p_{90} \qquad \text{Eq. 6}$$

We measured the pitch in four samples produced with a time interval $\tau = 30$ s but with $\theta_R$ equal to 5°, 11°, 45°, and 90°. The results are plotted in Fig. 3F along with the decreasing pitch as predicted from Eq. 5 & 6. The pitch of the structures follows the prediction and decreases from a few millimetres to a few hundreds of microns as the angle $\theta_R$ increases from 5° to 90°.

MASC-TGG therefore allows the creation of ceramics with periodically varying texture. The range of possible texture patterns is infinite as $\tau$ and the platelets alignment angle can be varied simultaneously. The control of the mold shape and topography opens the possibility to alter the texture not only in the z-direction but also in the (x-y) plane. With 10 μm-long grains, the ultimate smallest variation in the texture is thus also 10 μm, with virtually no limit on the largest variation.

Since the grain dimensions influence the mechanic properties of dense ceramics, monotextured samples should present anisotropic properties whereas the periodically-textured specimen should exhibit analogous oscillations in local mechanical response.

**Locally controlled mechanical properties**

To quantify the influence of the local variation of texture in dense ceramics on their local mechanical properties, we measured the Young's modulus and Vickers hardness of a monotextured MASC-TGG sample at different angles ψ, defined as the angle between the grain long axis and the testing direction, and compared these values with the periodic variation in hardness with the pitch of the periodic sample (Figure 4).

The Young's modulus, measured with an acoustic pulse-echo method, and the materials hardness, measured with a Vickers indenter at a load of 5 N, are plotted as a function of ψ in figure 4A. The Young's modulus follows a Voigt type rule of mixture:

$$E = \sqrt{E_0^2 \cos\psi^2 + E_{90}^2 \sin\psi^2}, \qquad \text{Eq. 7}$$

with $E_0$ = 120 +/- 20 GPa and $E_{90}$ = 25 +/- 10 GPa. The Young's modulus of a sapphire monocrystal is 10% higher along the crystallographic c-axis (415 GPa) than along the a or b-axis (385 GPa) (*35*, *36*). However, the difference between $E_0$ and $E_{90}$ follows an opposite trend, with the highest value obtained perpendicular to the c-axis. The difference cannot be explained by the intrinsic crystallographic anisotropy and is probably due to the presence of the 4.9 vol% of elongated pores and grain boundaries that also follows the texture obtained by TGG.

The sample's Vickers hardness similarly depends on the grain orientation and varies from $H_0$ = 11 +/- 4 GPa to $H_{90}$ = 3 +/- 1 GPa when ψ varies from 0° and 90°. This hardness follows a Reuss type of rule of mixture:

$$H = 1/\sqrt{\frac{\cos\psi^2}{H_0^2} + \frac{\cos\psi^2}{H_{90}^2}}, \qquad \text{Eq. 8}$$

which indicates that the softer damaging mechanism dictates the hardness of the sample. The larger number of grain boundaries and pores along the loading direction when ψ = 90° are probably providing an easy damaging mechanism. Polymer composites reinforced with alumina platelets exhibit similar trends with $H_0 > H_{90}$ and $E_0 > E_{90}$. For example, a polyurethane reinforced with 20 vol% of alumina platelets has mechanical properties of 0.25 and 0.19 GPa for $H_0$ and $H_{90}$ respectively, and 1.25 and 0.75 GPa for $E_0$ and $E_{90}$, respectively (*28*).

The measurement of monotextured samples proves that the mechanical properties of the ceramic depend on the testing direction with respect to the grain orientation. A periodically textured sample should thus present periodically varying mechanical properties as well. To demonstrate this, we performed *in situ* microindentation in an scanning electron microscope (SEM) with a Berkovitch indenter at a load of 1 N on a sample with a Bouligand structure of pitch 150 μm (Fig. 4B). Prior to indentation, the sample was polished via ion beam to obtain an almost atomically flat surface. However, the atomic re-deposition during the ion milling partially filled the pores, making the grain orientation identification more difficult. The microhardness values are indeed varying with the grain orientation, with values ranging from $H_0^{micro} = 18\ GPa$ to $H_{90}^{micro} = 8\ GPa$.

The microhardness values are 60% to 260% higher than the macrohardness ones, for $\psi = 0°$ and 90° respectively. This increase in material hardness as the indent imprint, thus the volume probed, is getting smaller is commonly observed in polycrystalline alumina (*37*). The size-dependence of the hardness is associated with a lower number of grain boundaries in the volume tested. The different shape of the indenter could also be responsible for a part of the difference between the macro and micro hardness. Nevertheless, the general trend with $H_0 > H_{90}$ is also observed. The oscillations in the hardness are visible as expected from the grain orientation angles variations. The SEM observation of the microindents further illustrates the significant role of the grain boundaries in activating plastic deformation mechanisms (Fig. 4C). The indentation performed with $\psi = 90°$ show more severe damages, with grain debonding and sliding, compare to the indentation performed at $\psi = 0°$. Cracks propagating at the interfaces between the grains are highlighted in yellow in both cases.

With the fabrication method developed, local variations in mechanical properties that directly depend upon grains' orientations can be incorporated into a 95 % dense alumina ceramic at the micrometer level. MASC-TGG therefore uniquely provides ceramics with locally tailored hardness or elastic modulus.

**Discussion**

The method presented here enables the creation of 95% dense alumina ceramics with a deliberate orientation of the constitutive grains by combining magnetically assisted slip-casting and templated grain growth. A composition compatible with TGG was used during MASC to fully enable the specificities from both techniques: the possibility to program local orientation of anisotropic particles in bulk 3D shapes from MASC and the ability to densify the assembly without pressure while producing highly aligned grains from TGG.

The alignment of the grains is almost as good as in ceramics prepared by tape casting, the state-of-the-art method, with a Full-Width at Half Maximum of the grain orientation distribution of 10°. But MASC-TGG is the only method that can achieve a similar control over arbitrary alignment directions.

The control of the local texture is translated into a local anisotropic mechanical response, both in elasticity and in hardness. Samples made with a periodically varying grain orientation present an analogous variation of the local hardness. These variations can be further modulated by adapting the grain size and aspect ratio through the sintering conditions and the addition of dopants.

To conclude, the combined use of MASC and TGG opens the possibility for locally programming the grain orientation in ceramic materials and thus to deliberately tune their local mechanical properties. These local variations of mechanical response are expected to translate into unusual macroscopic behavior as found in biological samples, through various phenomena such as crack arrest and twisting. With the increasing number of studies revealing the unique and exquisite microstructures of hard biological composites and minerals, often associated with outstanding properties, MASC-TGG finally provides tools to further explore how biological design principles could be transposed into ceramics. Further development of the process can be foreseen, such as the tailoring the final microstructure in terms of grain size and aspect ratio, density and local orientation. In addition, the scalability and the lesser efforts of the fabrication should be emphasized. Because it hinges on slip casting and pressure less sintering, two already industrial-scaled processes, MASC-TGG could be directly used for the fabrication in series of unusual ceramics. This local texturing can also be exploited beyond mechanics. Virtually any property that depends on the presence of a preferred orientation can be locally controlled with this technique and imply effective repercussions on the bulk macroscopic functionalities.

## Materials and Methods

**Magnetic functionalization of the template particles.** Anisotropic alumina platelets (Ronaflair White sapphire, Merck, Germany), used as the template particles for the grain growth, were magnetized following the procedure published elsewhere (*28*).

**Preparation of the suspension for casting.** The isotropic alumina nanopowder (TM-DAR, Taimei, Japan, average particle size 180 nm) was first dispersed in deionized water at 50 vol% solid loading and maintained at pH 4 (HCl 1 mM, Sigma Aldrich, Germany). Mechanical vibrations and addition of small amounts at each time were required to reach this high solid loading. After de-agglomerating overnight using ball-milling with 3- and 5-mm diameter alumina balls, magnetized platelets were added to the suspension to achieve the ratio nanoparticles:platelets of 90:10. Deionized water was added to adjust the total solid content back to 50 vol%. Before casting, the suspension was ultrasonicated (UP200S, Dr Hielscher) for a few minutes only at a power of 20% to ensure good dispersion, quickly degassed under vacuum, and kept under magnetic stirring until use. Due to the sensitivity to drying of highly concentrated suspensions, the recipient should be kept closed when not in used.

**Casting.** The porous molds were prepared in advance using gypsum (Boesner, Switzerland). These molds were cast and dried at 60°C at least one day. For cylindrical specimens, the gypsum was made flat by passing a razor blade at the surface. For specimen with a surface topography, the gypsum piece was scratched and cut until the desired surface was obtained. A plastic tube was then attached on top of the gypsum and maintained using parafilm foil to prevent leakage. This mold was then fixed on top of a stepper motor (Conrad, Switzerland) and positioned as close as possible to a neodymium magnet (Supermagnete, Switzerland) in an area where the magnetic field strength was at least 50 mT. The magnet was attached to a mixer that rotated at a frequency of 1 Hz at least. For periodically-textured specimen, the stepper motor was controlled by a LabView program sending a step function with tunable frequency and amplitude to tune the time $\tau$ and the angle step $\theta_R$. After complete removal of the water through the pores of the mold, which typically takes 3 hours, the mold with the sample was dryed overnight before unmolding.

**Sintering.** The dried samples were sintered in air in a furnace (Nabertherm, Switzerland) with a dwell time of 1 hour, a heating rate of 2.5 °C min$^{-1}$ with the cooling left unregulated. Except for the study of the grain growth, the sintering temperature used was 1600°C.

**Characterization of the microstructure.** The density of the sintered specimen was measured using a conventional Archimedes set-up in water. To ensure efficient water infiltration through the pores, vacuum was applied for a few hours. The grain sizes where determined in the samples with horizontal alignment. Cross-sections were polished, debris removed in an ultrasonic bath, coated with 5 nm of platinum and observed by scanning-electron microscopy (LEO 1530, Zeiss, Germany). Image analysis using Image J was used to measure the grain sizes over at least 100 grains per specimen. For the periodic specimen, brittle fracture of the cross-section were directly sputtered and visualized in SEM to reveal the local 3D orientation of the grains as shown in Figure 3E. Small angle X-ray scattering measurements were performed at the coherent beamline (cSAXS) at the Swiss Light Source (Paul Scherrer Institute, Switzerland). The X-ray wavelength of 0.1 nm was obtained from a double-crystal Si (111) monochromator and focused using the bendable monochromator crystal and high-order mirror. 2D diffraction patterns were recorded with the PILATUS 2M detector at a distance of 2.1489 m and an exposure time of 1s. The analysis of the data was performed using sasfit (https://kur.web.psi.ch/sans1/SANSSoft/sasfit.html).

**Determination of the misalignment.** Two complementary methods where employed to characterize the angle misalignment. First, fractured surfaces of monotextured aluminas were sputtered with platinum and observed in SEM. The Image J plugin Monogenic J (*38*) (http://bigwww.epfl.ch/demo/monogenic/) was used to assign a color to each dark edge in the images, in this case the grain interfaces. These colored images where then used as input into a Matlab program to associate each color with the relevant angle (details in the Supplementary Materials). For each orientation, at least 3 images at different locations in the specimen and containing a large number of grains (>100) were analysed and this for 2 samples. The same method was directly applied on SEM images available in the literature and on SEM images from a fracture surface of an Abalone shell. Second, X-ray rocking curves were recorded. They were obtained on polished samples with an Empyrean diffractometer (PANalytical, Germany) in reflection using the K$\alpha$ radiation from Cu at a wavelength of 0.15418 nm and around the (0012) crystallographic plane. The software TexturePlus was used to remove the beam divergence effect on the data.

**Measure of the pitch in function of the thickness.** The pitch of the various structures in function of the thickness was measured based on optical micrographs taken with incident light. The pictures where then converted to grey scale using Image J and direct measurement was done with the same software. The theoretical fits for the pitch were done using the equations 5 & 6.

**Mechanical characterisation.** Macroindentation was performed using a Vickers indenter (MXT-a, Wolpert, Germany). The values were averaged over 10 measurements. The elastic modulus was measured using the ultrasonic method in pulse echo in reflection at 5

MHz (longitudinal probe V1091 5MHz, shear probe V157-RM 5MHz, couplant SWC-Shear Wave Couplant, 072PR-20-E, General Pulser Receiver, Controltech, Switzerland) on discs of 3-mm thickness $L$ and polished to ensure flat surface. The time-of-flight *tof* was recorded using an oscilloscope. The Young's modulus was calculated using $E = \rho V_t^2 \frac{3V_l^2 - 4V_t^2}{V_l^2 - V_t^2}$, with $\rho$ the density, and $V_l$ and $V_t$ the longitudinal and transverse wave velocity, respectively, and $V = \frac{2L}{tof}$. *In-situ* micro indentation was performed using the Alemnis SEM indenter (Alemnis, Switzerland) with a Berkovitch tip at 1N force and using a LabView-controlled position. The ceramic sample was initially fixed on the SEM stub and polished using broad-ion-beam (Hitachi IM4000, Japan) for 2.5 hours (6kV, deviation angle C3, Argon) to ensure a very flat surface.

## Acknowledgments

**General**: We acknowledge Marianne Liebi, Ofer Hirsch, Sebastian Krödel and Jeffrey Wheeler for their help with the experiments and discussion and Tobias Niebel for providing electron micrographs of natural nacre. We thank the microscopy center of ETHZ, ScopeM, for access to the broad ion beam facilities. **Funding:** The work was supported by internal funding from ETH Zürich and from the Swiss National Science Foundation (grant 200020_146509). **Author contributions:** H.L.F and F.B. designed the

research, H.L.F performed the experiments and H.L.F. and F.B. prepared the manuscript. H.L.F, F.B. and A.R.S discussed the results and their implications and revised the manuscript.

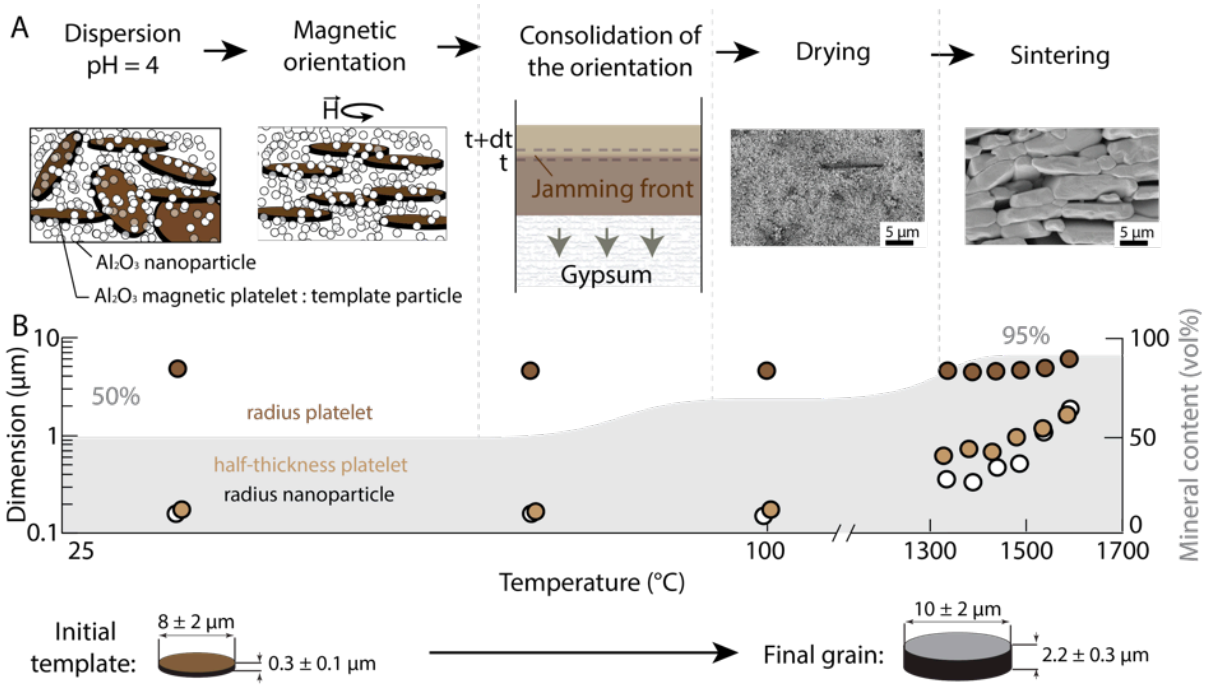

**Fig. 1. Pressure-less densification of alumina ceramics with magnetically-controlled texture orientation**. **A)** Schematics and electron micrographs at the different stages of the process, namely initial dispersion of magnetic alumina template particles and alumina nanoparticles in acidic conditions, magnetic orientation of the templates, consolidation of their orientation by solvent removal through the pores of a porous mold, solvent evaporation and after template thermal grain growth by sintering at 1600°C. The last two micrographs are cross-sections of fractured surfaces. **B)** Evolution of the dimensions of the platelets and nanoparticles with the temperature during the entire process along with the mineral content (in grey). The schematics highlight the growth of the initial template particles into large anisotropic grains.

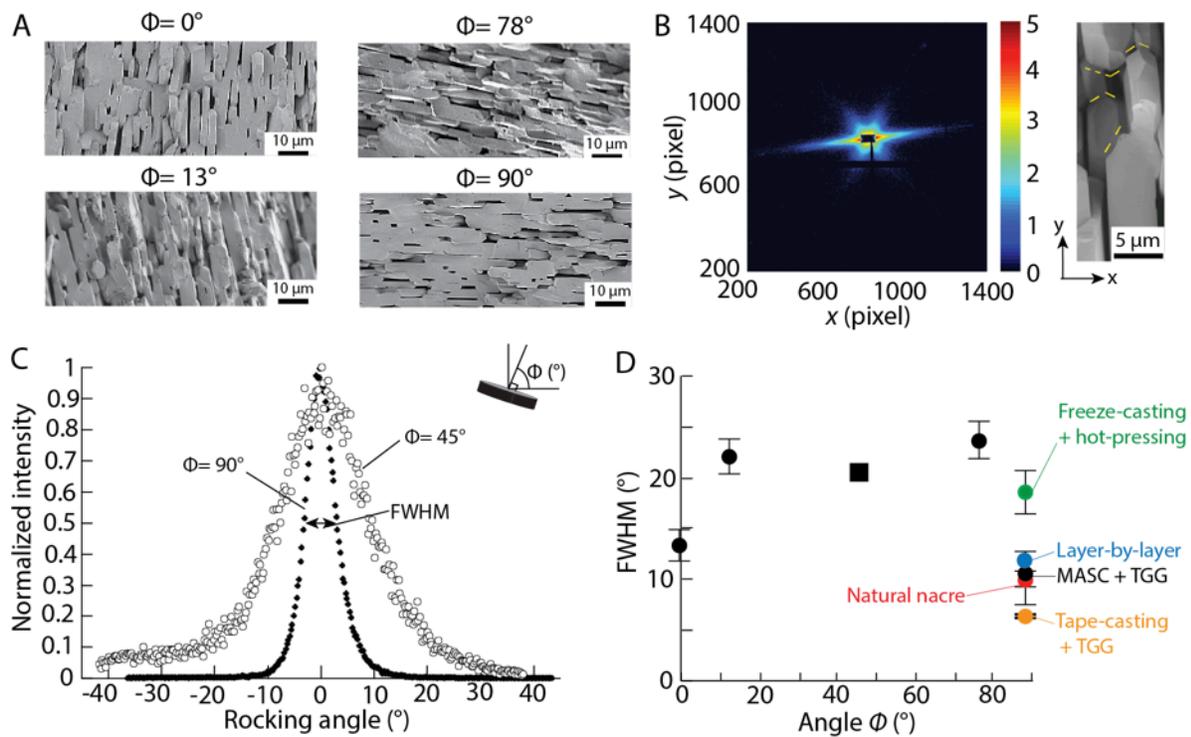

**Fig. 2. Texture control along arbitrary directions. A)** Electron micrographs of fractured surfaces of dense monotextured alumina with various texture directions, defined by the angle $\phi$ (°) representing the angle of the normal of the grain with the horizontal. **B)** Small-Angle X-ray scattering pattern of a monotextured sample with $\phi=0°$ as in the electron micrograph where the short edges of the grains are highlighted in yellow. **C)** Rocking curve of monotextured specimen with $\phi=90°$ and with $\phi=45°$. **D)** Grain misalignment using the full-width at half maximum (FWHM) as shown in (**C**) for monotextured dense ceramics as prepared in this study (MASC-TGG, black circle measured by image analysis and black square with rocking curve), using tape-casting (*11*), layer-by-layer process (*32*) or freeze casting (*13*) and compared with the natural nacre from an Abalone seashell (see Fig. S2).

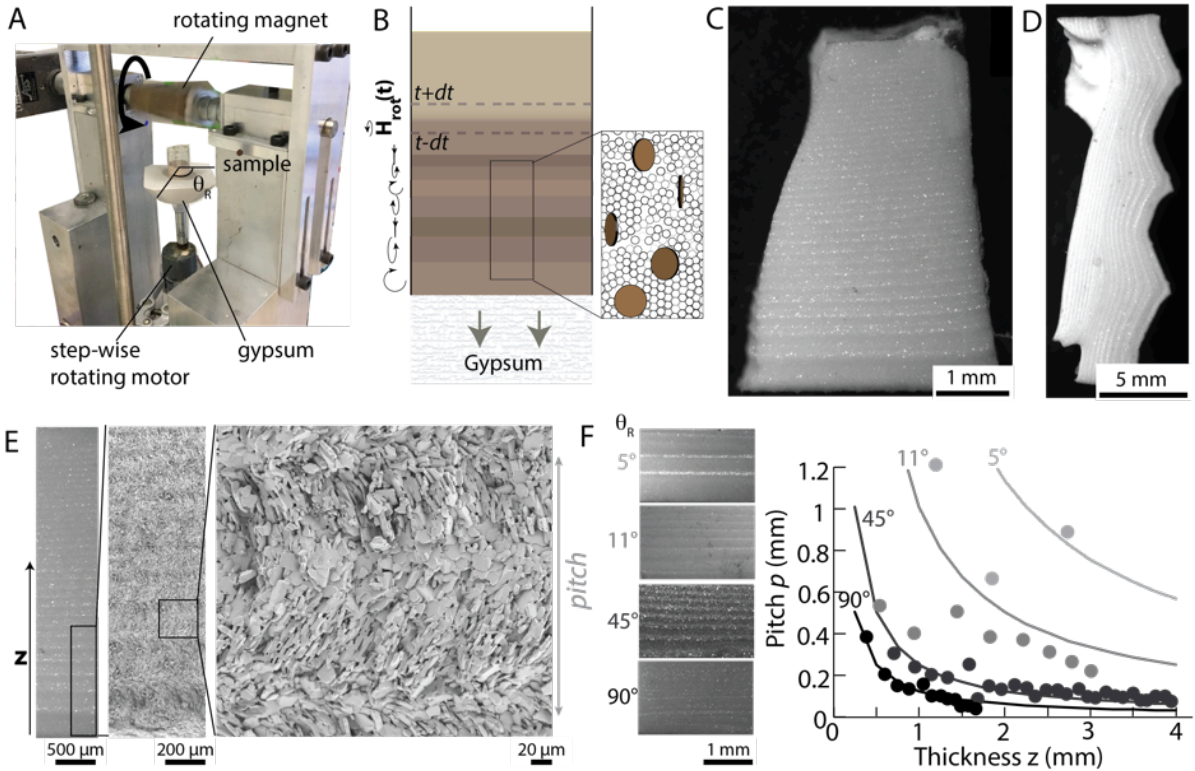

**Figure 3: Fabrication and control of the pitch of periodically textured ceramics by MASC-TGG. A)** Setup used to obtain a periodic ceramic with vertical orientation of grains. The step rotating motor turns the sample by an angle $\theta_R$ at each step while a rotating permanent magnet aligns the template particles vertically. **B)** Schematics of layer-by-layer texture pile-up during the growth of the jamming front during MASC. Optical micrographs of the final dense ceramic featuring the periodic layering **(C)** and the conformation of the periodic structure on the sample's surface topology **(D). E)** Electron micrographs showing the periodic microstructure of the grain orientation and the definition of the pitch on a sample made with a $\theta_R = 90°$ and a time interval $\tau = 30$ s. **F)** Optical measurement of the influence of the angle $\theta_R$ imposed during the casting on the pitch of the final structure, lines have been fitted according to equations 5 & 6.

Page **21** of **24**

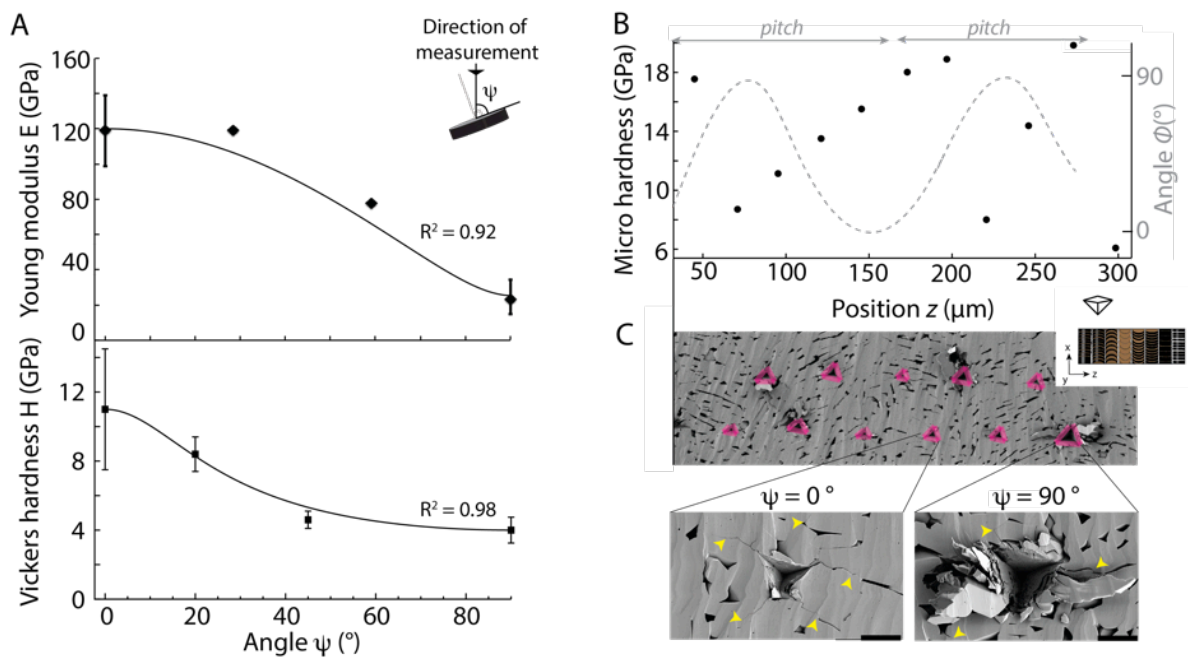

**Figure 4: Texture-dependent mechanical properties of monotextured and periodically textured samples prepared by MASC-TGG. A)** Dependence of the Young's modulus E (top) and Vickers macro hardness H (bottom), performed at a load of 5 N, as a function of the angle ψ between the testing direction and the grains' long axis direction. **B)** Micro-indentation series at 1 N along a pitch of a periodically textured MASC-TGG. The angles of the grains are estimated according to the pitch and measured optically by reflection. **C)** The electron micrographs show the series of indents (pink) along the pitch of the periodically varying microstructure placed as in the insert, with close view on the different types of damage (yellow arrows) made at an angle of ψ = 0° and ψ = 90°.

**Supplementary Materials**

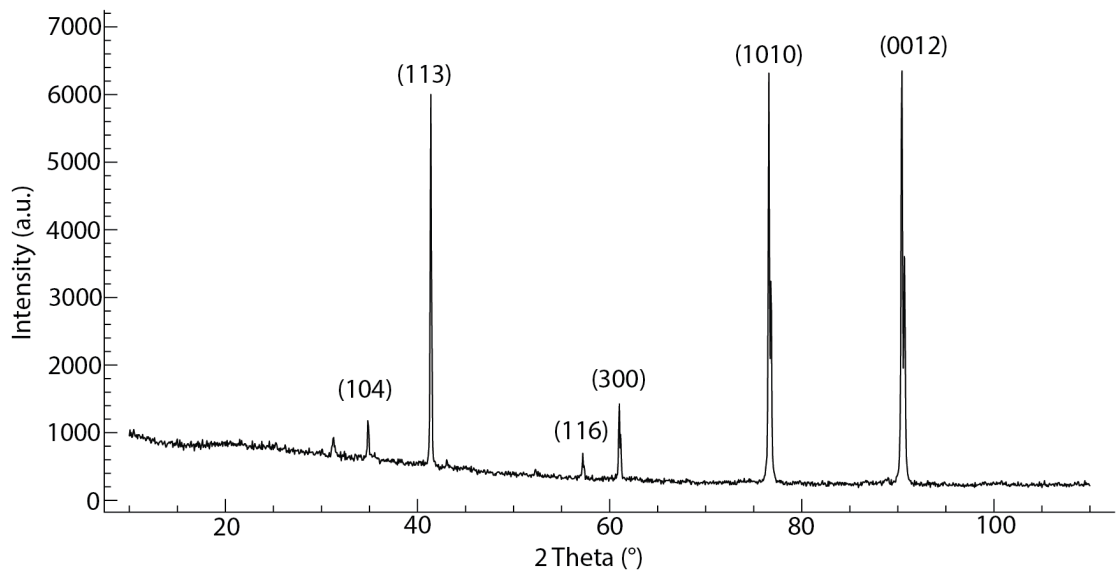

**Figure S1: X ray scattering of an MASC-TGG alumina ceramic with horizontal alignment ($\phi = 0°$).**

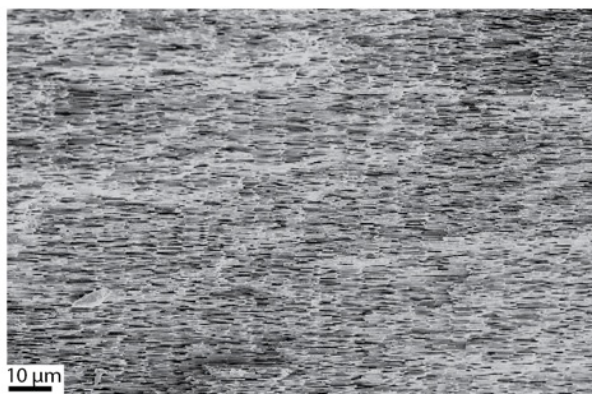

**Figure S2: Electron micrograph of a fractured surface of an Abalone seashell.**

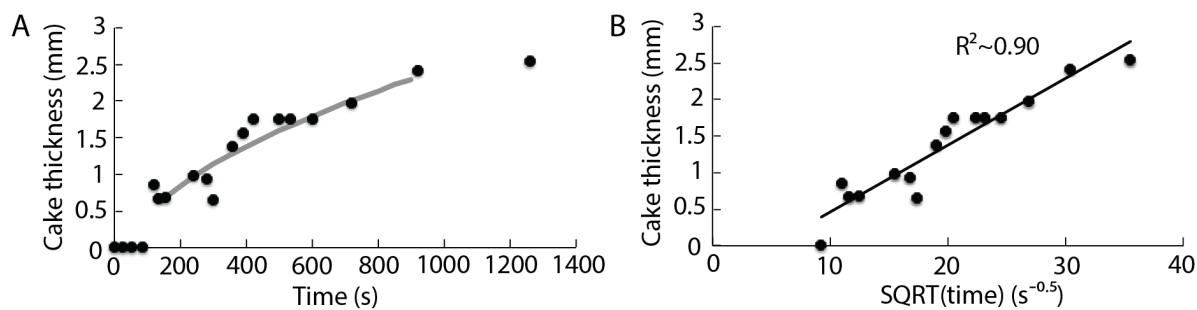

**Figure S3: Kinetics of the jamming front during MASC. (A)** the thickness of the jammed deposit, traditionally called cake thickness, from the surface of the porous mold in function of the casting time (black dots) and the fit of the experimental data according to equation 1 (grey line). **(B)** Cake thickness in function of the square root of the time used to determine the kinetics growth parameter A by linear regression.

**Matlab code used after the plugin Monogenic to obtain the interfaces angles distribution:**

```matlab
% Open image RGB stack from MonogenicJ, named here "Orientation2"
    im = Orientation2;
    [sy sx] = size(im) ;
    totallength = sy*sx;

    % transform the matrix into one single line
    vect = reshape(im, [1,totallength]);

    %tabulate : gives in the first column the value, in second column the
    %count and in the third the percentage

    tbl = tabulate(vect);
    long = 237;
    %always put 255 before

     % plots the distribution (value in degree - percentage)
    for k=1:long
       value(k)=(tbl(k,1)*pi/long)-pi/2;
       valued(k)=value(k)*180/pi;
       per(k)=tbl(k,3);
    end

   plot(valued,per)
f = fit(valued.',per.','gauss2')
plot(f,valued,per)
```